\documentclass[preprint]{elsarticle}
\newcommand{\degree}{\ensuremath{^\circ}}
\journal{Astronomy \& Computing}

\begin{document}

\begin{frontmatter}

\title{CADRE: The CArma Data REduction pipeline}

\author[dnf]{D. N. Friedel}

\address[dnf]{University of Illinois, Department of Astronomy, 1002 W. Green St., Urbana, IL 61801}

\begin{abstract}
The Combined Array for Millimeter-wave Astronomy (CARMA) data reduction pipeline (CADRE) has been developed to give investigators a first look at a fully reduced set of their data. It runs automatically on all data produced by the telescope as they arrive in the CARMA data archive. CADRE is written in Python and uses Python wrappers for MIRIAD subroutines for direct access to the data. It goes through the typical reduction procedures for radio telescope array data and produces a set of continuum and spectral line maps in both MIRIAD and FITS format. CADRE has been in production for nearly two years and this paper presents the current capabilities and planned development.
\end{abstract}

\begin{keyword}
data reduction pipeline
\end{keyword}

\end{frontmatter}

\section{Introduction}
The Combined Array for Millimeter-wave Astronomy (CARMA) radio telescope array is located in the Inyo mountains in California, USA. The array is heterogeneous and is composed of 23 antennas (six 10 m dishes, nine 6.1 m dishes, and eight 3.5 m dishes). The array observes in the 1mm, 3mm, and 1cm wavelength regimes, in both continuum and spectral line modes, and has dual polarization capabilities at 1mm. The array can observe in two independent subarrays, observing different sources at different frequencies with different correlator configurations at the same time. All data produces by the telescope are in the MIRIAD \citep{miriad,miriadascl} format.

The primary purpose of CADRE \citep{cadreascl} is to give the investigators a quick first look at a fully reduced version of their data. This reduced data is of the quality produced by the average user. CADRE runs automatically on all CARMA data as it arrives in the CARMA archive.

CADRE is capable of reducing the following observing modes: continuum, spectral line, mixed mode (spectral and continuum in the same track), single pointings, multipoint mosaics, and Sunyaev–Zel'dovich effect. Future upgrades of CADRE will be able to reduce dual polarization, full stokes, and CARMA23 (all 23 antennas observing together) data.

\section{The Software}\label{sec:software}
The back end of CADRE is written primarily in Python ($\sim$24K lines of code), using calls to MIRIAD routines to do the majority of the data processing. In order to reduce the reliance on fragile flat text files produced by many of the MIRIAD tasks, Python wrappers for the MIRIAD subroutines were developed. These wrappers were developed with SWIG\footnote{http://www.swig.org} to directly wrap the Fortran and C subroutines into python callable methods. During development it was discovered that errors encountered by the subroutines caused issues. The underlying subroutines do not raise exceptions but throw segmentation faults, which in turn would crash out the Python session. In order to prevent Python from completely crashing mxTools\footnote{http://www.egenix.com/products/python/mxBase/mxTools/} were added to the wrappers. These tools catch the segmentation faults and translate them into Python exceptions which can then be caught and dealt with appropriately. These wrappers are publicly available in the CVS distribution of the MIRIAD package (\$MIRSRC/scripts/python/subwrap) as of version 4.1.6, but currently need to be manually built. Additional routines were developed for CADRE for specific tasks (e.g. reading a MIRIAD image into a numpy array for easy calculations and manipulation). These additional tasks are dependent on a few publicly available 3rd party Python packages: numpy\footnote{http://www.numpy.org} and threadpool\footnote{http://chrisarndt.de/projects/threadpool/}.

In order to optimize CADRE for speed, many of the pipeline tasks which operate on individual spectral widows (i.e. gain calibration, map creation, cleaning) are threaded at the Python level. The time it takes a data set (track) to run through CADRE is highly dependent on the array configuration and number of pointings. In general it takes a track anywhere from a few minutes to many hours to process, with the median being less than 30 minutes. CADRE was turned on 20 Sept.\ 2011 and is successful (producing full output) on more than 95\%\footnote{This does not include unsupported modes and tracks with failing grades.} of datasets. The remaining 5\% fail for a variety of reasons, most of which are failures of the MIRIAD tasks to converge on a proper solution. The data of any failures is returned to the end user in the log files (see\S~\ref{sec:output}).

The full pipeline codebase is available for download from a public archive at github.com/astro-friedel/CADRE

\section{The Process}
The following sections describe the process and logic CADRE uses while working with the data. CADRE uses a user preferences file to control some of the processing. Table~1 gives the parameters that can be adjusted in the preference file.

\begin{center}
Table 1: Preference File Description
\end{center}
\begin{tabular}{lp{6cm}r}
\label{tab:pref}
Parameter & Description & Default \\ 
\hline
tsysThreshold & Flag data with a system temperature above the given threshold & 5000.0 \\ 
BIMAShadowFraction & Shadowing fraction for 6.1 m dishes (i.e. if this fraction of the radius is shadowed then flag it) & 0.8 \\ 
OVROShadowFraction & Shadowing fraction for 10.4 m dishes (i.e. if this fraction of the radius is shadowed then flag it) & 1.0 \\ 
SZAShadowFraction & Shadowing fraction for 3.5 m dishes (i.e. if this fraction of the radius is shadowed then flag it) & 1.0 \\ 
doBaselines & Apply the appropriate baseline solution & True \\ 
doDecorrelation & Do decorrelation correction & False \\ 
selfcalInterval & Time interval for selfcalibration (minutes) & 5.0 \\ 
bootfluxInterval & Time interval for \verb#bootflux# averaging (minutes) & 5.0 \\ 
amplitudeGainRange & If the amplitude gains of an antenna are outside of this range then flag the associated data & [0.2,5.0] \\ 
maxAmplitudeGainFactor & If the mean gain of an antenna is more than this factor above the general mean, then the antenna is flagged & 3.0 \\ 
maxGainRms & If the gain rms of an antenna is above this value, then the antenna is flagged & 1.0 \\ 
cellSize & The cell size used to invert the data (in arcsec), the default value lets CADRE decide based on the average baseline length & -1.0 \\ 
imageSize & The image size to invert (in pixels), the default value lets CADRE decide based on the cell size and primary beam size & -1.0 \\ 
cleanThreshold & How deep should clean iterate when compared to the rms of the map (in units of rms noise) & 5 \\ 
cleanRegion & What region should the clean algorithm concentrate on, values can be any valid MIRIAD region selection command & inner quarter \\ 
doContinuumSubtraction & Should CADRE attempt continuum subtraction & False \\ 
doAutoCleanRegion & Should CADRE attempt to determine the clean region automatically & False \\ 
\hline
\end{tabular} 

Before the formal data reduction begins CADRE scans the MIRIAD file in order to identify and classify all the sources (passband calibrator, gain calibrator, flux calibrator, source, etc.), determine the observing frequency, window parameters (bandwidth, channel width, frequency), observation date, etc.

\subsection{$u-v$ Data Calibration}
Calibration of the raw {\it u-v} data goes through the following steps (Note that all flagging is done with the MIRIAD task \verb#uvflag#, unless otherwise noted.):

\subsubsection*{Antenna Positions}
All CARMA baseline solutions are stored in the MIRIAD CVS tree. If the doBaselines option in the preferences file is set to {\it True} (the default) then CADRE searches for the most recent baseline solution in the appropriate array configuration and applies it to the data using the MIRIAD task \verb#uvedit#. It is only recommended to set this parameter to {\it False} if the antenna positions in the data set are known to be good, in this instance the antenna positions already in the data are used.

\subsubsection*{Flagging Base on System Temperature}
Any data with system temperatures above the threshold given in the preferences file are flagged.

\subsubsection*{Flagged Based on Antenna Shadowing}
The MIRIAD task \verb#csflag# flags data based on antenna shadowing. The task uses the values from the preferences file to set shadowing thresholds for the different antenna types. Additionally the task will use a swept volume calculation to determine if any antennas were shadowed by another antenna not in the same subarray. The calculation looks at what volume of space each antenna could be occupying and if any of this space could shadow another antenna it is assumed that it does and flags the data appropriately.

\subsubsection*{Flagging based on Elevation}
CADRE will flag any data taken at high elevation ($>$87\degree) as the antennas can have difficulty tracking at these elevations.

\subsubsection*{Passband Calibration}
\begin{figure}[ht!]
\includegraphics[height=\textwidth ,angle=270]{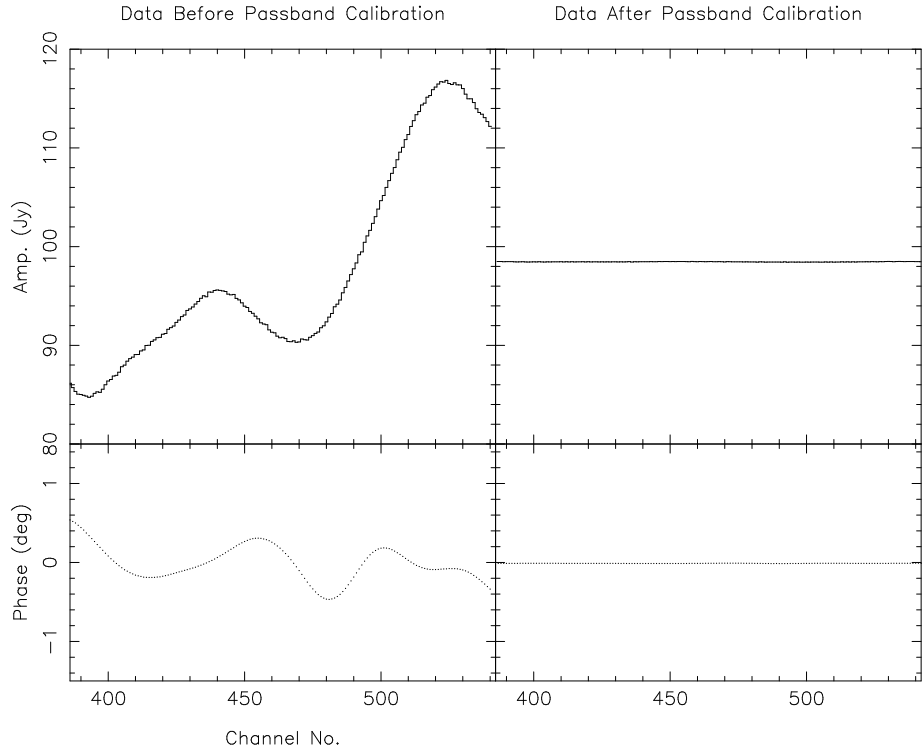}
\caption{An example of a good passband solution. The left hand panels are before calibration (amplitude and phase), while the right hand panels are after calibration (amplitude and phase). \label{fig:passcal}}
\end{figure}

\begin{figure}[ht!]
\includegraphics[height=\textwidth ,angle=270]{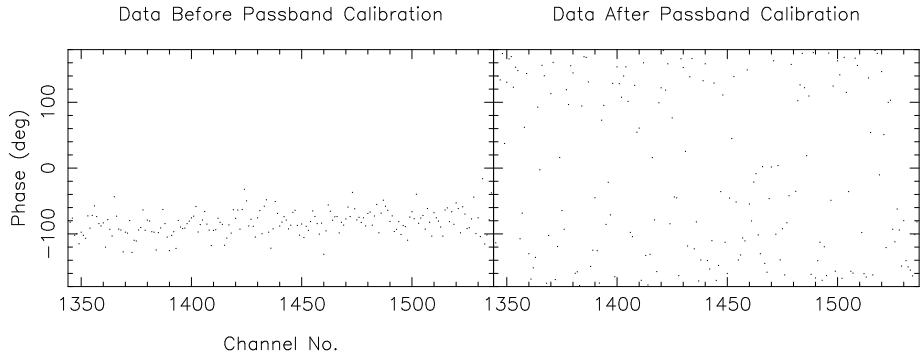}
\caption{An example of a bad passband solution. The left hand panel is before calibration, while the right hand panel is after calibration. Note that the phase scatter went up notably. This solution would cause the offending antenna(s) to be flagged by the system and the solution to be recalculated. \label{fig:badpasscal}}
\end{figure}

CADRE will correct each spectral window for passband effects (frequency dependent artifacts induced on the data by the sky and instrument). There are several ways that the passband can be corrected, and the best method depends on the spectral resolution of each window. For the channel widths above $\sim$1 MHz the preferred method is to use an astronomical source (quasar or planet). If no appropriate calibrator was observed then CADRE will fall back to the internal noise source. For the narrower channel widths only the internal noise source has enough signal to noise to generate a useful solution. The passband corrections are calculated with the MIRIAD task \verb#mfcal#. Figure~\ref{fig:passcal} shows an example of a good passband solution. The left hand panels are before calibration (amplitude and phase), while the right hand panels are after calibration (amplitude and phase). Before applying the passband solutions to the data CADRE calculates the rms noise of the phase solution for each window. If the rms is greater than 50\degree\ \footnote{An rms of this level usually indicates a bad solution or bad data, thus to be on the safe side CADRE will flag the associated data.} then the associated window is flagged and the passband solution is recalculated. Figure~\ref{fig:badpasscal} shows what a bad passband solution looks like.

\subsection*{Decorrelation Correction}
CADRE defines decorrelation to be the increased scatter of amplitudes with increasing baseline length, due to atmospheric turbulence. To correct for decorrelation CADRE first determines whether the data are decorrelated, based on the gain calibrator data. This is only an issue for the longest baselines in the most extended array configurations. If the data are less than 25\% decorrelated, no attempt at correction is made. Otherwise CADRE attempts to correct for the atmospheric decorrelation with the the MIRIAD task \verb#uvdecor#. CADRE then determines the decorrelation of the "corrected" data. If the new decorrelation is better than the previous value then the correction is applied to the rest of the data. If it is worse then CADRE reverts to the previous data and no correction is made. For long baseline data the correction can increase the rms noise, however the corrected data more correctly reflect that actual source structure on those Fourier size scales.

\subsubsection*{Flagging of Bad Amplitudes}
\begin{figure}[ht!]
\includegraphics[height=\textwidth ,angle=270]{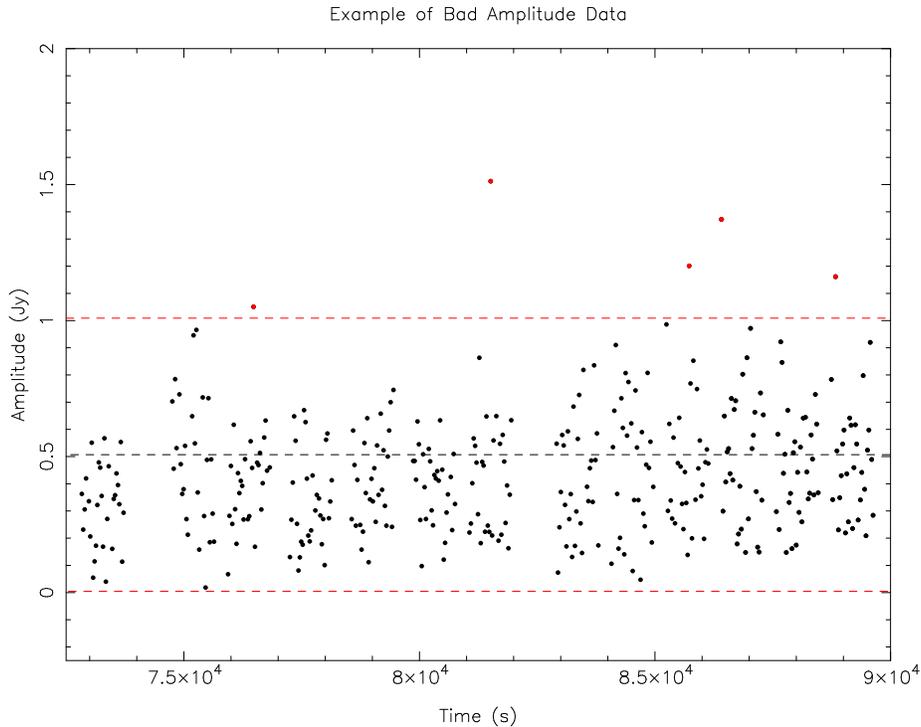}
\caption{An example of a bad amplitude points (noted in red) that would be flagged by CADRE. The black dashed line denotes the mean value, while the red dashed lines denote $\pm$2 times the rms noise.\label{fig:badSrcAmp}}
\end{figure}

High and low amplitude anomalies on source and calibrator data are flagged (based on the rms and mean values). Any individual integration that is more than 2 times the rms above or below the mean amplitude is flagged. Figure~\ref{fig:badSrcAmp} shows an example of bad amplitude points, noted in red, that would be flagged by CADRE.

\subsubsection*{Flagging Spectral Birdies}
CADRE attempts to detect spectral birdies and flag the affected data. The system defines a spectral birdie as a
strong, narrow (single channel) signal in both sidebands: it must be in the same channel number in both
the upper and lower sidebands. Birdies are extremely rare in 1mm and 3mm CARMA data, but can cause calibration
issues if present. There are known birdies in the 1cm system, which are avoided during observations.

\subsubsection*{Amplitude Calibration}
CADRE attempts to determine the absolute amplitude of the gain calibrator(s) in the data. There are several methods of doing this. They are described here in decreasing preference:
\begin{enumerate}
\item If a planet was observed during the track CADRE runs the MIRIAD task \verb#bootflux# on both the planet and gain calibrator(s) to calculate the amplitude. The bootfluxInterval parameter in the preferences file is used to control the length of time over which \verb#bootflux# calculates each interval, typically the time which is spent on each calibration cycle.
\item If no planet was observed, or \verb#bootflux# fails, CADRE searches the data for a secondary calibrator (e.g. a strong quasar used for passband calibration) and runs \verb#bootflux# on it and the gain calibrator(s) to determine the amplitude.
\item If this fails then CADRE looks at the internal MIRIAD calibrator flux tables, which are regularly updated with dedicated flux monitoring observations.
\end{enumerate}

\subsubsection*{Gain Calibration}
\begin{figure}[ht!]
\includegraphics[width=0.8\textwidth]{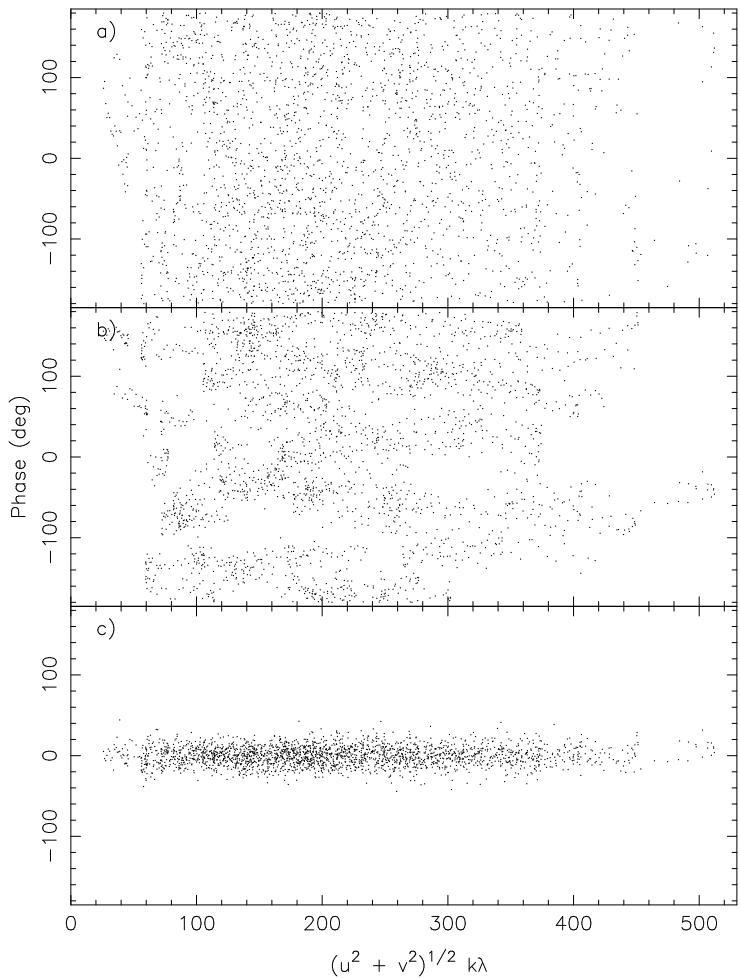}
\caption{An example of bootstrapping the gain solution from a wideband window to a narrower one. a) shows the uncalibrated narrowband data. b) shows the same data after application of the gains from the wideband window, note that there is some coherence but still a lot of scatter. c) shows the same data after a phase only selfcalibration with a long averaging interval is applied, these data are now very coherent.  \label{fig:bootstrap}}
\end{figure}

CADRE attempts to be as cautious as possible when it computes the amplitude and phase gains from the calibrator(s). The MIRIAD task \verb#mselfcal# is used to calculated the gains. If no absolute flux was determined by the amplitude calibration routines then only the phase gains are calculated, otherwise both amplitude and phase gains are calculated. The selfcalInterval parameter in the preferences file is used to control the averaging interval for the selfcalibration solutions, typically the time which is spent on each calibration cycle. Multiple gain calibrators are each handled independently, and the solutions are applied in succession to the data. In the case of homogeneous bandwidths (e.g. all 500 MHz windows) then all data in each sideband are used together in the solutions. In the case of heterogeneous bandwidth configurations (e.g. a mix of 500 MHz and 32 MHz bands), which is typical for spectral line observations, the widest bandwidth windows in each sideband are used to calculate the solutions. This solution is then copied to the narrower bandwidth data and a second \verb#mselfcal# run is completed on these data. This solution is done in phase only and with a long averaging time (typically hours) in order to remove the window based phase offsets. Figure~\ref{fig:bootstrap} shows an example of bootstrapping the gain solution from a wideband window to a narrower one. a) shows the uncalibrated narrowband data. b) shows the same data after application of the gains from the wideband window, note that there is some coherence but still a lot of scatter. c) shows the same data after a phase only selfcalibration with a long averaging interval is applied, these data are now very coherent.

\subsubsection*{Flagging Errant Gains}
CADRE will flag calibrator data based on gains computed by the task \verb#mselfcal#. CADRE looks for anomalously high or low amplitude gains (cutoff values are set in the user preferences file with the keyword amplitudeGainRange) and flags the data appropriately. Additionally, CADRE will compare the average amplitude gains of each antenna and flag any antenna whose gains are more than a given number of times that of the others (specified with maxAmplitudeGainFactor in the preferences file). The typical cause of this is a bad pointing solution for an individual antenna. Additionally, CADRE looks at the rms of the amplitude gain solutions for each antenna, and if any are above 1.0 (as specified with the keyword maxGainRms in the preferences file) the associated antenna is flagged. If any flagging is done the selfcalibration solutions are recalculated before proceeding.

\subsubsection*{Application to Source Data}
The gains from the calibrator(s) are copied and applied to the source data with \verb#gpcopy# and \verb#uvcat#. In the case of heterogeneous bandwidths the gain solutions are copied and applied incrementally with each selfcalibration solution. In the case of multiple gain calibrators CADRE will apply the gains from each calibrator in succession, making sure that only 1 solution is applied for each time interval.

\subsubsection*{Flag Unbracketed Source Data}
\begin{figure}[ht!]
\includegraphics[height=0.8\textwidth ,angle=270]{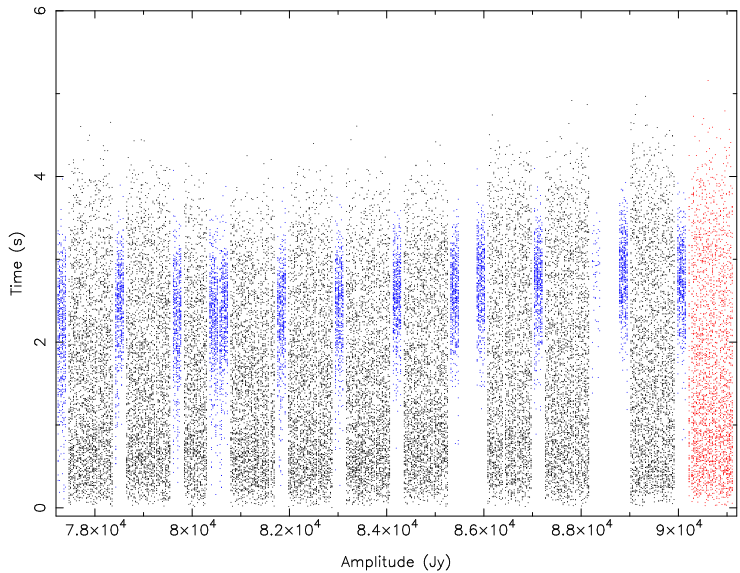}
\caption{An example of unbracketed source data. The calibrator data are shown in blue and the bracketed source data are in black. The unbracketed (not surrounded by calibrator data in time) source data are shown in red and will be flagged by CADRE. \label{fig:unbracketed}}
\end{figure}

In the rare instance of a track ending early and the last calibration cycle is not observed, then CADRE will flag any source data that is not surrounded by gain calibrator data (in time). Figure~\ref{fig:unbracketed} shows an example of unbracketed source data. The calibrator data are shown in blue and the bracketed source data are in black. The unbracketed (not surrounded by calibrator data in time) source data are shown in red and will be flagged by CADRE.

\subsection{Mapping}
CADRE will produce several mapping products: a continuum map for each source, spectral line maps for each window for each source, and short and long baseline continuum maps for Sunyaev–Zel'dovich effect observations, for point source subtraction. The following sections give the details of the process.

\subsubsection*{Inverting}
CADRE uses the MIRIAD task \verb#invert# to generate the dirty map and dirty beam. Before this task is run, several parameters need to be calculated, namely the cell size to use and the image size to produce. If the cell size is not specified in the preferences file (keyword cellSize) then CADRE calculates it based on the estimated synthesized beam size. CADRE aims to get between 3 and 5 pixels across the synthesized beam while choosing a reasonable value (i.e. 0.2" and not 0.2265"). The image size, unless specified in the preferences file (keyword imageSize), is based on the FWHM of the largest primary beam and the cell size, specifically the FWHM in arcseconds divided by the cell size (also in arcseconds) yields the image size in pixels.

All continuum maps are generated with the \verb#options=systemp,mosaic,mfs# specified. These options specify weighting the visibilities by their system temperatures, treat the observations as a mosaic, and do multi-frequency synthesis maps (properly construct the map given the input visibilities span a wide range of frequency) respectively. The maps are treated as a mosaic due to the heterogeneous nature of the telescopes. This ensures that all the data get the proper weights based on antenna/priamry beam type. The mosaic option is not necessary for small, point-like sources near the phase center, but as CADRE does not have any apriori knowledge of the source structure it was felt that this option should be specified in all instances. Spectral line cubes are generated with just \verb#options=systemp,mosaic# .

\subsubsection*{Cleaning}
The cleaning process is an iterative one. The MIRIAD task \verb#mossdi# is used to do the deconvolution. The default is to clean the inner quarter of all single pointing observations and the FWHM regions of the 10m antennas of multi-point mosaics. This can be overridden in the preferences file (keyword cleanRegion). CADRE can also be commanded to determine its own clean region (using the keyword doAutoCleanRegion in the preferences file). It does this by reading the output of the first clean/restore cycle (see below) into an numpy array and searching for emission above 5 times the rms noise level. Any located emission regions are stored and at the end of the search a clean region is calculated based on these emission regions.

The first run of \verb#mossdi# is a very shallow clean ($\sim$100 iterations per channel). Next the MIRIAD task \verb#restor# is run on the deconvolved image and beam to produce a cleaned image. CADRE then calculates the rms noise of the map. These statistics are used to set the cutoff values for subsequent runs of \verb#mossdi#. The cutoff is typically set to 5 times the rms noise of the map, this is set by the cleanThreshold parameter in the preferences file. The clean/restore loop is continued as long as the rms noise decreases after each iteration, or if 5 iterations are reached. In the case of spectral line observations the clean/restore loops of the individual windows are threaded in Python, which greatly increases the speed of the reduction process.

\subsubsection*{Continuum Subtraction}
CADRE can attempt to do continuum subtraction for the spectral line windows (keyword doContinuumSubtraction). Each spectral window is searched for spectral lines. This is done by scanning the data cube for spectral emission peaks. Using these peaks, channels containing spectral lines are identified and removed from the list of line free channels. Once all spectral line channels have been removed from the list the MIRIAD task \verb#uvlin# is used to subtract the continuum, using the line free channels as a model. After the continuum is subtracted the spectral windows are re-reduced to produce continuum free spectral data cubes.

\subsection{Future Development}
There are several capabilities under development for CADRE:
\begin{itemize}
\item Ability to reduce dual-polarization data
\item Ability to reduce full stokes data
\item Ability to reduce CARMA23 mode data (all 23 antennas observing the same target with multiple correlators)
\item Ability to reduce Maximum Sensitivity mode data (multiple correlators attached to a single subarray)
\item Ability to reduce multiple tracks with the same target source and combine them into one map
\end{itemize}

\section{Output}\label{sec:output}
\begin{figure}
\includegraphics[width=0.5\textwidth]{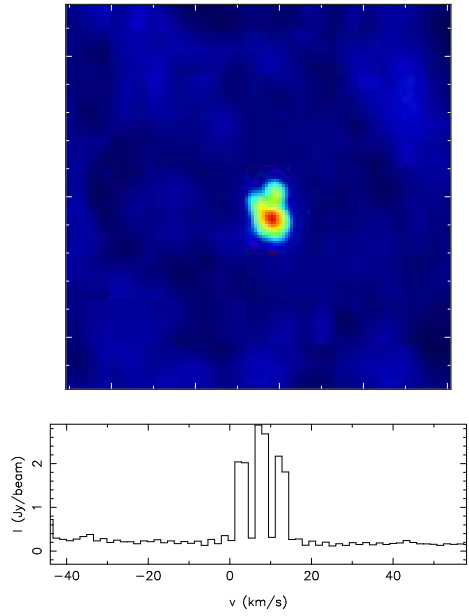}
\caption{Spectra and channel map of dimethyl ether produced by CADRE.\label{fig:spec}}
\end{figure}

\begin{figure}
\includegraphics[height=0.5\textwidth ,angle=270]{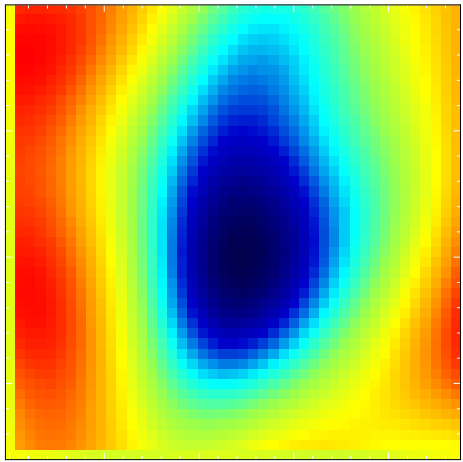}
\caption{Sunyaev–Zel'dovich decrement map produced by CADRE.\label{fig:decrement}}
\end{figure}

\begin{figure}
\includegraphics[height=0.5\textwidth ,angle=270]{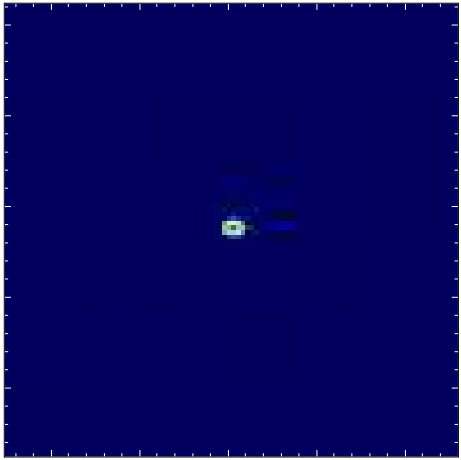}
\caption{Gain calibrator map produced by CADRE. The calibrator map can be used to check the quality of calibration.\label{fig:cal}}
\end{figure}

CADRE provides a wide array of output data files to the end user. These include:
\begin{itemize}
\item calibrated {\it u-v} data for all sources and calibrators
\item a continuum map of the primary gain calibrator
\item continuum maps of all source
\item spectral line maps of all spectral windows, if spectral line mode was specified
\item a short baseline continuum map for Sunyaev–Zel'dovich effect observations
\item a long baseline continuum map for Sunyaev–Zel'dovich effect observations (for point source subtraction)
\item a csh script that contains all of the MIRIAD commands used to reduce the data, this script can be run on the user's home machine to exactly reproduce CADRE's work, or it can be tweaked by the user and re-run to alter the data reduction (this assumes the user has MIRIAD installed)
\item a notes file that gives more detail on what was done to reduce the data, this includes comments on what processing was being done and any errors encountered
\item a copy of the current CADRE files so the user can run CADRE on this or other data on their home machine\footnote{Running CADRE on the user's home system does require the installation of some 3rd party software, in addition to MIRIAD (version 4.1.6 or newer), see \S\ref{sec:software} for a detailed list of what is needed. It is always recommended to have the most recent version of MIRIAD running.}
\end{itemize}
Each output map is produced in both MIRIAD and FITS format (using the MIRIAD task \verb#fits#). Figure~\ref{fig:spec} shows a plane from a channel map produced by CADRE. It shows a map of dimethyl ether and the associated spectra. Figure~\ref{fig:decrement} shows a Sunyaev–Zel'dovich effect decrement map produced by CADRE. Figure~\ref{fig:cal} shows a gain calibrator map produced by CADRE. Calibrator maps can be used to quickly check the quality of the calibration process, as a single point source at the center should be all that exists in the map. All output files are tar'd together and stored in the CARMA data archive\footnote{Accessed at http://carma-server.astro.illinois.edu:8181} for retrieval by the PI along with the raw data. All CADRE data products are subject to the same proprietary access restrictions as the raw data (18 months for non-thesis data, from the date of observation).

\section{Summary}
This paper has introduced the CArma Data REduction pipeline (CADRE) and gives a detailed description of its processes. CADRE goes through typical data reduction procedures for radio telescope array data and produces maps to give the investigator(s) a first look at the quality calibratability of their data. All CADRE products are stored in and retrievable from the CARMA data archive, with the same proprietary access restrictions as the raw CARMA data. CADRE will continue to be developed to handle more of CARMA's current and future capabilities.

\subsection*{Acknowledgements} The author would like to thank the numerous individuals who beta tested the software and offered suggestions for improvement. Support for CARMA construction was derived from the Gordon and Betty Moore Foundation, the Kenneth T. and Eileen L. Norris Foundation, the James S. McDonnell Foundation, the Associates of the California Institute of Technology, the University of Chicago, the states of California, Illinois, and Maryland, and the National Science Foundation. Ongoing CARMA development and operations are supported by the National Science Foundation under a cooperative agreement, and by the CARMA partner universities.

\bibliographystyle{model2-names}
\bibliography{pipeline}

\end{document}